# Text-based Modeling

Hans Grönniger, Holger Krahn, Bernhard Rumpe, Martin Schindler and
Steven Völkel

Institute for Software Systems Engineering
Technische Universität Braunschweig, Braunschweig, Germany
http://www.sse-tubs.de

**Abstract.** As modeling becomes a crucial activity in software development the question may be asked whether currently used graphical representations are the best option to model systems efficiently. This position paper discusses the advantages of text-based modeling over commonly used graphical representations. It is inspired through the advent of new extensible development tools like Eclipse. The discussion is illustrated by showing a textual version of UML state machines as Eclipse plugins.

## 1 Text-based Modeling

Modeling becomes an increasingly important technique for the development of complex software systems. Nowadays, three different approaches are used for modeling: textual languages, graphical languages and combinations of text and graphic. The latter usually being dominated by graphics, where text is just a supplement. Currently, the UML is the most common visual approach for modeling, as it is both widely known and assisted by many tools developed in the last decade.

Most of them do not allow to conveniently look at the textual parts. E.g. if only one method body is visible at a time in a pop-up window, then efficient comparison etc. is not possible. Therefore practical work with these tools often leads to manipulation of generated code and hence enforces a round-trip-approach. Experienced developers argue that a text-based approach with conventional editors is much more convenient.

To assist an efficient agile, yet model-based development process, we have developed a framework called MontiCore [12] in the last two years, which supports text-based modeling. In contrast to generic approaches like [15] an arbitrary concrete syntax can be used for a modeling language which increases its usability. MontiCore offers several textual languages for different purposes like a textual version of UML/P [19, 20] including OCL, languages for feature modeling, and a language for architectural description. This paper discusses the advantages of text-based modeling based on the experiences we made so far. These experiences indicate that there are indeed some advantages over the currently dominating graphical-based approach. These advantages are partially permanent and partially will disappear, when the tooling (editors, visualization, incremental code generation, context checks on models) will be considerably improved. However, based on the progress tool vendors made in recent years, it will probably take at least another decade, until we will actually see such tools that are visual, efficient and convenient.

In Section 2 we explain advantages of text-based modeling for the language user. As existing tool support enhances the users productivity we list arguments in Section 3 how tools can be created easily for textual languages. Section 4 compares our position with others and lists examples for textual modeling languages. In Section 5 a textual modeling language is defined and supporting tool development is sketched as an example how such a language may be developed and used. Section 6 concludes the paper.

## 2 Advantages for the Language User

The advantages we identified so far, differ concerning the roles that participate in a development process [11]. From the developers viewpoint, the advantages are:

**Information content.** Graphical models usually need more space than text-based models to display the same information content. So while it is generally easier to get an overview of a graphical model compared to a textual model, the graphical model is more wide-spread and thus less information can be seen on a screen or a sheet of paper. To understand the details, either many pop-ups are necessary or a lot of scrolling/zooming has to be done. As developers usually are more concerned with details than with grasping a first understanding of a model, text-based models seem to have advantages in the long run.

Furthermore, text can be printed very easily, while graphical models, especially for complex and huge systems, often exceed the size of a sheet of paper in horizontal and vertical direction. Experiences also show that the information on one page can more easily be grasped from graphical documents, but the higher number of printed pages nullifies this advantage.

**Speed of creation.** With today's tools writing text is a lot more efficient than drawing graphical models. The latter often use drag-and-drop, menus and pop-ups to fill in details of the graphically depicted form. While this is very helpful for the inexperienced modeler, working with these graphical tools is a time-consuming task. Modelers constantly have to switch between mouse and keyboard. Only one pop-up is visible at a time and it takes a number of mouse-clicks to switch between different elements. In addition these menus and wizards tend to enforce complete definitions before allowing the modeler to progress adding data elsewhere. This seriously hampers the creative development process and hinders evolution.

**Integration of languages.** Not everything that needs to be described during a development process should be depicted graphically. Visual languages [21] tried to do so and failed because of lack of developer efficiency. Today, it is common among modeling tools that some parts, in particular for describing conditions and actions should be textual. However, these are often badly integrated or not conveniently reachable. If all languages are textual, the integration is much easier and leads to more understandable models.

**Speed and quality of formatting.** The placement of graphical models is a time-consuming process which distracts the user from the actual modeling task. This could be done by layout algorithms, but even for small models their results are often not acceptable. Formatting text is a much easier task and the results of standard algorithms are of high quality. The main reason is that depending on the semantics of the model elements the developer has a special layout in mind whereas automatic algorithms cannot consider this information.

**Platform and tool independency.** Since text requires neither a specific platform nor a specific environment for reading and modifying, the development, enhancement and correction of models can be done almost anywhere. Our experiences show that this is an important advantage over graphical languages because it is possible to read and modify the models in different environments without installing additional tools. This advantage even holds, if we provide special support (syntax highlighting, etc.) in form of an Eclipse plugin, because Eclipse runs on many platforms and even more important, its use is not mandatory.

**Version control.** Version control plays an important role in software development teams. Today's most commonly used version control systems, CVS and SVN are text-based and therefore can easily be used for an efficient version control for text-based models as well. Both can be included in nearly every IDE, and more important, both can be used on command line to preserve platform and tool independency.

Although there are approaches for visual models to store them in repositories, neither is the calculation of differences between these models fully understood yet, nor do merging algorithms of those models work in general. However, merging of text on line basis works very well. Thus, for textual models normal text-based version control systems like CVS or SVN can be used. Experiences show, that using such version control systems on serialized models in XMI format does not work either, because little changes in the visual models can lead to large changes in the respective serialization and thus to conflicts. Furthermore, due to the unreadable XML data format conflicts are not easily resolvable by users.

Using text-based models also has some disadvantages for the developer. E.g., as discussed, graphics are more intuitive in order to get a first orientation. Modern text-based development tools have incorporated this idea by usually giving an outline of the detailed code, e.g., in form of a list of defined elements. This outline can also be a tree, showing defined states or used methods of a state machine for example.

It is of course a big advantage of visual tools that simulation and animation look a lot more intuitive using graphics. This might be an argument to provide a graphical representation for simulation purposes even, if just a textual language is used for development.

However, one may argue that the best solution would be an interchangeable format giving the user the choice between graphics and text. In this case, the abovementioned points such as lower speed and quality of formatting and especially the tool dependency are still valid.

## 3 Advantages for the Tool Developer

Text-based modeling, however, does not only have advantages for the developer (which is the user of the modeling tool), but also for the developer of a modeling tool. These advantages become particularly interesting, when the forthcoming multitude of domain specific languages (DSLs, [2]) is considered. The many new DSLs that will emerge must easily be definable together with appropriate tool support. Using textual languages has a number of advantages here as well:

**Editors (almost) for free.** A textual language can be handled by using ordinary textual editors. Even, if syntax high-lighting and auto-completion are of interest, this can be achieved rather easily in a number of editing environments.

**Outlines and graphical overviews.** As CASE tools show, it is possible to create graphical views from textual input. Sophisticated graphical modeling tools like MetaEdit+ [14] simplify the creation of new modeling languages. However, it still takes a lot of effort to use these tools and the experiences with our MontiCore framework or similar approaches like ASF+SDF [22], TCS [10] and Xtext [16] demonstrate that support for developing text-based languages can still be a lot more efficient.

**Parsers, pretty print, code generators and translators are rather easily developed.** Other standard tools, like a parser is developed as easily as an XML-parser using appropriate tools, like ANTLR [17] or a DSL-definition framework like MontiCore [12]. The latter e.g. allows to develop internal representation of the abstract syntax (meta-model) according to the given concrete textual form of the model. Definition of good layout-rules as well as development of a pretty printer are much easier.

**Composition of modeling languages.** Many domain specific languages will be similar to each other, will be extensions of given programming or modeling languages or would like to reuse subsets of other languages. For example ArchJava [1] or LINQ [13] use these techniques to improve the usability of programming languages.
An easy, reuse-enabling composition of modeling languages in various forms is therefore inevitable and can to our experience much easier be achieved by using textual languages. Here, syntactic and tool integration can be realized by shared symbol tables and attribute grammars. In MontiCore we developed an infrastructure for composing languages as well as their lexical analyzers and parsers allowing a first composition of languages on the tooling level. This language composition technique increases the reuse of existing languages and therefore, simplifies the tool development.

## 4 Empirical Studies

Papers which present graphical modeling tools usually do not compare their approaches to textual versions. They implicitly assume that "graphical representations are better simply because they are graphical" which is questioned in [18]. Results of a case study in which concrete problems are modeled using textual and graphical notations are analyzed. The authors argue that both text and graphics have their limitations and quality is not achieved automatically, although the authors reason that graphics have a higher potential of misleading the reader.

Finding further empirical evidence in literature to support or invalidate our hypotheses that textual modeling is underrated and superior to graphical modeling in a lot of cases is difficult. To support our view, we list a non-representative list of textual modeling languages that are preferred although graphical notations and tools exist. In [9] the development of a textual meta-modeling language is described because existing graphical tools lack usability. The authors state that after two years of experimenting with their approach they are convinced of the practicality. The tool USE [4] that supports validation and animation of UML specifications uses class diagrams and OCL invariants as inputs in textual form. USE outputs sequence and object diagrams graphically. Similarly, Alloy [8] for modeling and analyzing object systems consists of a textual specification language. Additionally, a graphical representation is provided. Another example for the usage of textual syntax for modeling languages is defined in the ITU-TS recommendation Z.120 [7]. It introduces a grammar for Message Sequence Charts used to describe interactions between objects or processes [6]. Furthermore, [5] indicates that visual programs of a language for circuit design were harder to comprehend than corresponding textual programs.

## 5 Text-based Modeling using Eclipse

To demonstrate the validity of the positions discussed above, we briefly sketch our text-based modeling approach and its advantages. The approach used is given in form of generated Eclipse plugins as nowadays integrated development environments (IDE) are more popular among developers than plain text editors. Especially Eclipse gained tremendous attention due to its extensible nature [3].

Eclipse allows the development of language specific functionalities. To our experience this extension mechanism is powerful, but it requires quite a lot of (unnecessary) repetitive programming to add assistance for any new textual language. This is an ideal target for code generation for MontiCore that allows the generation of syntax-driven editors, which support syntax-highlighting, outlines, and foldable source parts. In addition to language specific plugins, Eclipse supports the user by providing language independent services like version control including a conflict resolving editor and build management tools.

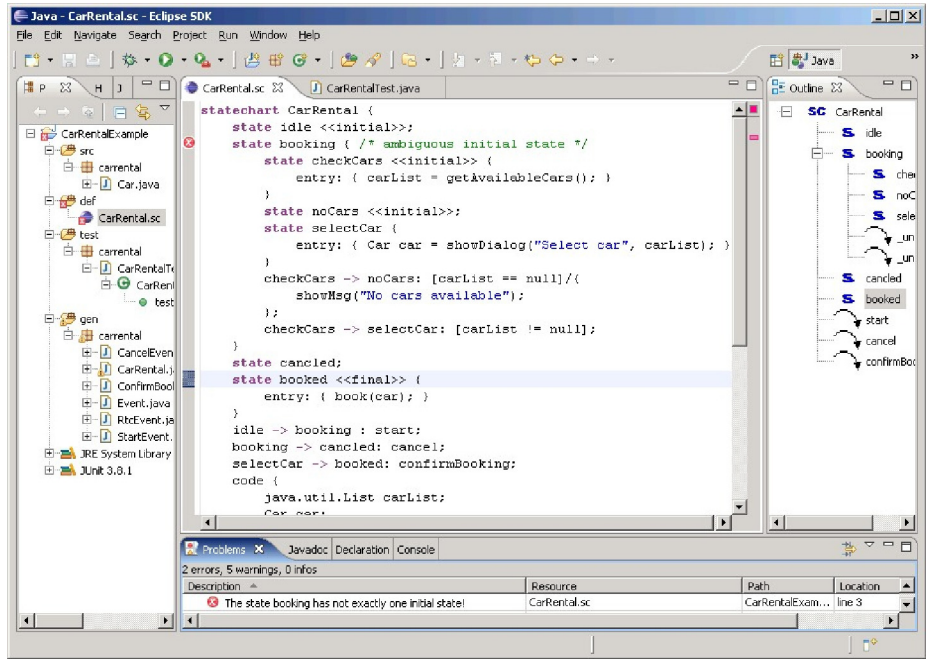

**Fig. 1.** State machine defining a simple car rental system

### Example

We demonstrate the advantages of text-based modeling by a textual version of the UML state machines [20]. Our state machines embody hierarchical states, but no parallel states or history, and allow the use of Java expressions as state invariants and preconditions of transitions. Additional methods and attributes can be defined outside of the state machines. In this form they can be integrated in any Java project as an executable modeling language (Figure 1).

From the language user point of view this version of state machines is easily applicable. To ensure platform and tool independency the plain text files defining a state machine can be modified using a simple text editor. Because of the intuitive and simple syntax used for the state machines concurrent changes reported by a version control system are often solved automatically. This is in contrast to graphics, where even simple changes like the redefinition of a transition can cause complex reformatting in the graphical representation of a state machine and a useful representation of both versions of the changes are hard to visualize. The formatting of the textual state machine is done by fast and simple indentation instead of time-consuming adjustment of graphical elements. Modifications can be done without any pop-ups or switching between mouse and keyboard. For a more comfortable editing we generated an Eclipse plugin which includes different comfort functions.

To reuse already existing languages as much as possible, the definition of the Java expression language is used within the state machine language. By combining both parsers for the state machine and Java, the code can be processed and syntactically checked before the program code is generated. In this way syntax errors in both languages are shown in the state machine itself and not in the generated code as common in other CASE-tools. Thus errors are more transparent for the language user.

The development of this modeling language was highly supported by MontiCore offering the definition and processing of textual languages including language composition. As a grammar for Java was already included in MontiCore only the state machine language had to be defined. Additionally, most of the infrastructure described above could be generated using MontiCore.

All this is convenient and efficient: The tool developer has considerably reduced efforts to make such a modeling language working, especially when composed from existing fragments of other languages. The system developer can efficiently use the editors and in particular the model analyzer and code generator, which allows a much more agile form of software development.

## 6 Conclusion

In this position paper we discussed arguments for text-based modeling and described first experiences that we made, when developing and using text-based modeling languages. There is some evidence, that text-based modeling constitutes a noteworthy alternative to graphical modeling because of its simple usage, scalability and easy development and reuse of tool support. This becomes particularly interesting when many domain specific languages will be developed in the next decades.

To substantiate the described experiences we are planning to set up case studies where we compare a textual version of the UML/P [19, 20] to other mainly graphical UML modeling tools with respect to handling and understanding models of different style and size, as well as the resulting effects in terms of efficiency and quality of the developed system.

*Acknowledgment:* The work presented in this paper is undertaken as a part of the MODELPLEX project. MODELPLEX is a project co-funded by the European Commission under the "Information Society Technologies" Sixth Framework Programme (2002-2006). Information included in this document reflects only the authors' views. The European Community is not liable for any use that may be made of the information contained herein.